\documentclass{moriond}

\def\be{\begin{equation}}
\def\ee{\end{equation}}
\def\bea{\begin{eqnarray}}
\def\eea{\end{eqnarray}}

\newcommand{\Photo}{}

\begin{document}
\vspace*{4cm}
\title{TOP QUARK PROPERTIES AT ATLAS AND CMS}

\author{ Jan van der Linden }

\address{Karlsruhe Institute of Technology,\\ 
	Kaiserstr.~12, 76131 Karlsruhe, Germany}

\maketitle\abstracts{
The latest results of the ATLAS and CMS experiments on top quark property measurements and interpretations in the EFT framework are presented. All measurements are based on the full LHC Run 2 dataset of approximately $138\,\textrm{fb}^{-1}$ taken at $\sqrt{s} = 13\,\textrm{TeV}$.}

\section{Introduction}

With the large amount of data taken during the Run 2 of the LHC, properties of top quarks can be measured with high precision.
This dataset of approximately $138\,\textrm{fb}^{-1}$ is utilized by the CMS and ATLAS experiments for various precision measurements of top quark properties. Measurements are performed by the CMS collaboration of CP violation~\cite{cpv} in the production and decay of top quarks and by the ATLAS collaboration of the energy asymmetry~\cite{ttj} of top quarks and anti quarks and the top (anti) quark polarization~\cite{polarization} in single top t-channel production.
Interpretations of the energy asymmetry~\cite{ttj} and the top (anti) quark polarization~\cite{polarization} measurements are performed in the effective field theory (EFT) framework.
EFTs are used as a framework to parameterize non Standard Model (SM) contributions to the known SM field theory ($\mathcal{L}_{\textrm{SM}}$) at higher orders in an effective Lagrangian. The EFT corresponds to an approximation of non-SM physics at a high energy scale $\Lambda$ to the (in comparison) low energies accessible at the current colliders. 
Operators $O^{d}_{i}$ of dimension $d$ are added to the effective Lagrangian $\mathcal{L}_{\textrm{eff}}$ as
\begin{equation}
\mathcal{L}_{\textrm{eff}} = \mathcal{L}_{\textrm{SM}} + \sum_{i} \frac{c_{i}}{\Lambda^{d-4}}O^{d}_{i}~,
\end{equation}
where $c_{i}$ are coefficients representing the coupling strength of the corresponding operators, suppressed by the scale $\Lambda^{d-4}$. This factor suppresses operators at higher orders, which leaves dimension-6 operators as the prime interest of current EFT interpretations. 
Some dimension-6 operators and the associated coefficients $c_{i}$, usually represented in the Wilson basis, can modify couplings of top (anti) quarks and thereby affect the measurements of top quark properties.
With the high precision achieved in these top quark property measurements, small amounts of hypothetical non-SM contributions modifying the top quark couplings could lead to detectable differences in the measured quantities.

The large amount of data also facilitates high precision measurements of differential cross sections of top quark anti quark pairs ($t\bar{t}$) also in association with additional radiation of e.g.\ a heavy boson. Small amounts of non-SM contributions modifying top quark couplings can affect such differential distributions, opening up the possibilities to interpret the differential measurements in the EFT framework due to the high precision of the differential measurements.
Two differential measurements are performed by the ATLAS collaboration~\cite{ttsl,ttfh}, targeting $t\bar{t}$ events in the boosted regime, which are interpreted in the EFT framework.
At the CMS collaboration limits on anomalous top quark couplings are set using $t\bar{t}$ events produced in association with a prompt photon~\cite{ttgammasl,ttgammadilep} and Higgs or Z bosons~\cite{ttZH}.
An additional measurement by the CMS collaboration~\cite{tZEFT} directly targets EFT coupling parameters in single top or $t\bar{t}$ events produced in association with a Z boson.

\section{CP Violation}

A measurement is performed by the CMS collaboration~\cite{cpv} to search for CP violation in the production and decay of top quarks.
For that purpose, four linearly independent observables $O_3, O_6, O_{12}$ and $O_{14}$ are constructed.
These four observables are triple products of spin- and momentum vectors of the final state particles expected in $t\bar{t}$ decays in the $1\ell$ channel.

To achieve sensitivty on CP violating effects, these operators are chosen such that they are odd in time reversal. 
Under CPT invariance, asymmetries 
\begin{equation}
A_{CP}(O_{i}) = \frac{ N_{\textrm{events}}(O_i > 0) - N_{\textrm{events}}(O_i < 0)}{N_{\textrm{events}}(O_i > 0) - N_{\textrm{events}}(O_i < 0)}~,
\label{eq:cpvasym}
\end{equation}
 in these observables hint towards CP violating contributions.
The measurement yields $A_{CP}$ values compatible with zero for all four observables $O_i$ with a precision of $0.1\%$. This constitutes an improvement by a factor of three relative to a previous measurement with data taken by CMS at $8\,\textrm{TeV}$~\cite{oldcpv}.

\section{Top quark anti quark asymmetries}
Asymmetries between top quarks and anti quarks in the production of $t\bar{t}$ pairs are measured by the ATLAS collaboration~\cite{ttj} targeting $t\bar{t}$ pairs produced in association with another jet ($t\bar{t}{+}j$).
The asymmetries are measured as differences in the energy of the top quarks and anti quarks.
As the asymmetry primarily originates in quark-gluon or quark-quark induced $t\bar{t}$ production modes, the boosted regime at high energies, which favors these production modes, is used.
This measurement is performed differentially in the jet scattering angle $\theta_j$ of the additionally radiated jet.
A high sensitivity in this measurement is achieved by defining the differential jet scattering angle based on the boost direction $y_{ttj}$ of the full $t\bar{t}{+}j$ system.
The result of this measurement confirms the SM prediction, that the top anti quark is on average produced at higher energies than the top quark, at around $3\%$, depending on the jet scattering angle $\theta_j$.

\begin{figure}
	\begin{minipage}{0.45\linewidth}
		\centerline{\includegraphics[width=0.99\linewidth]{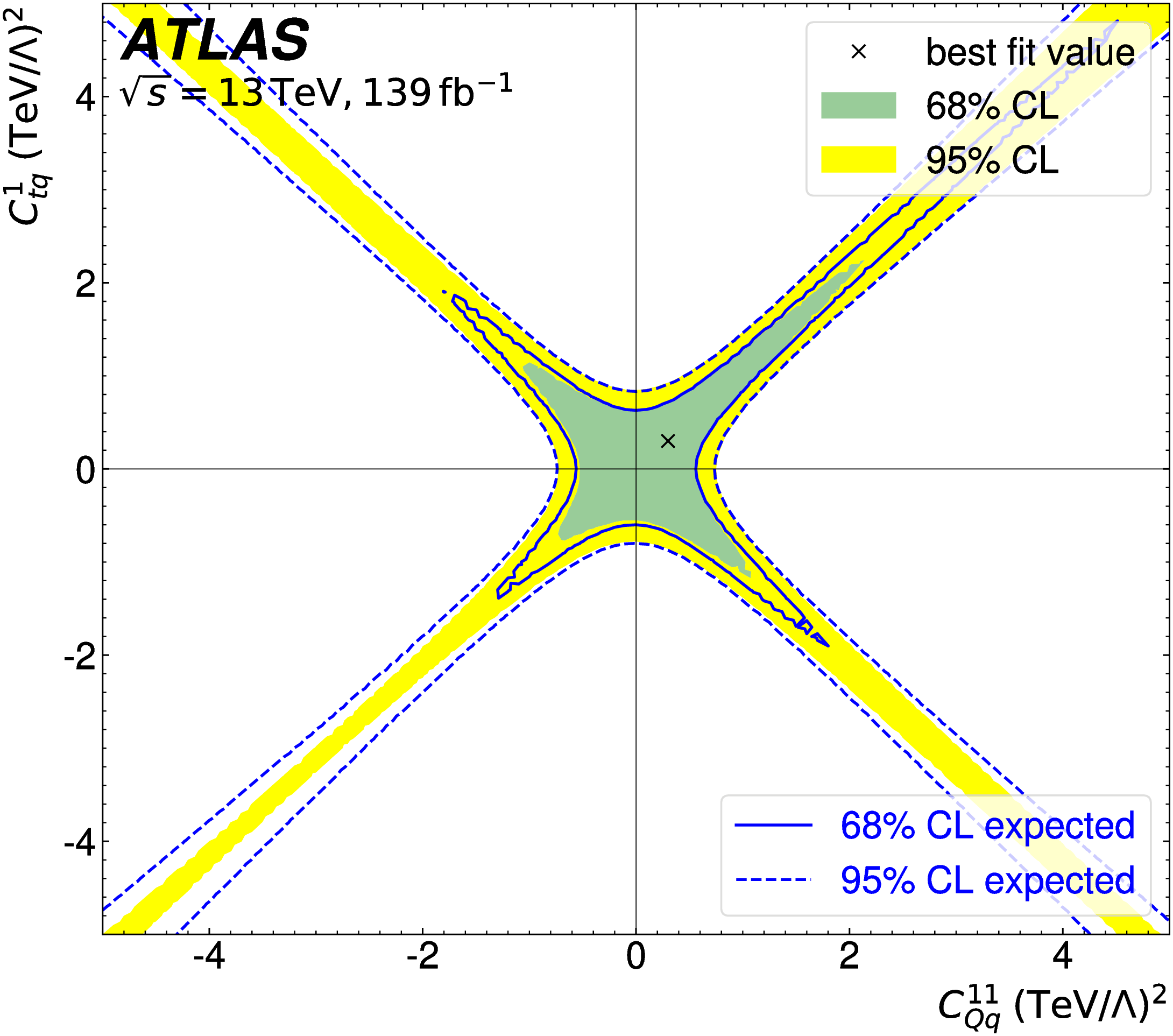}}
	\end{minipage}
	\hfill
	\begin{minipage}{0.45\linewidth}
		\centerline{\includegraphics[width=0.99\linewidth]{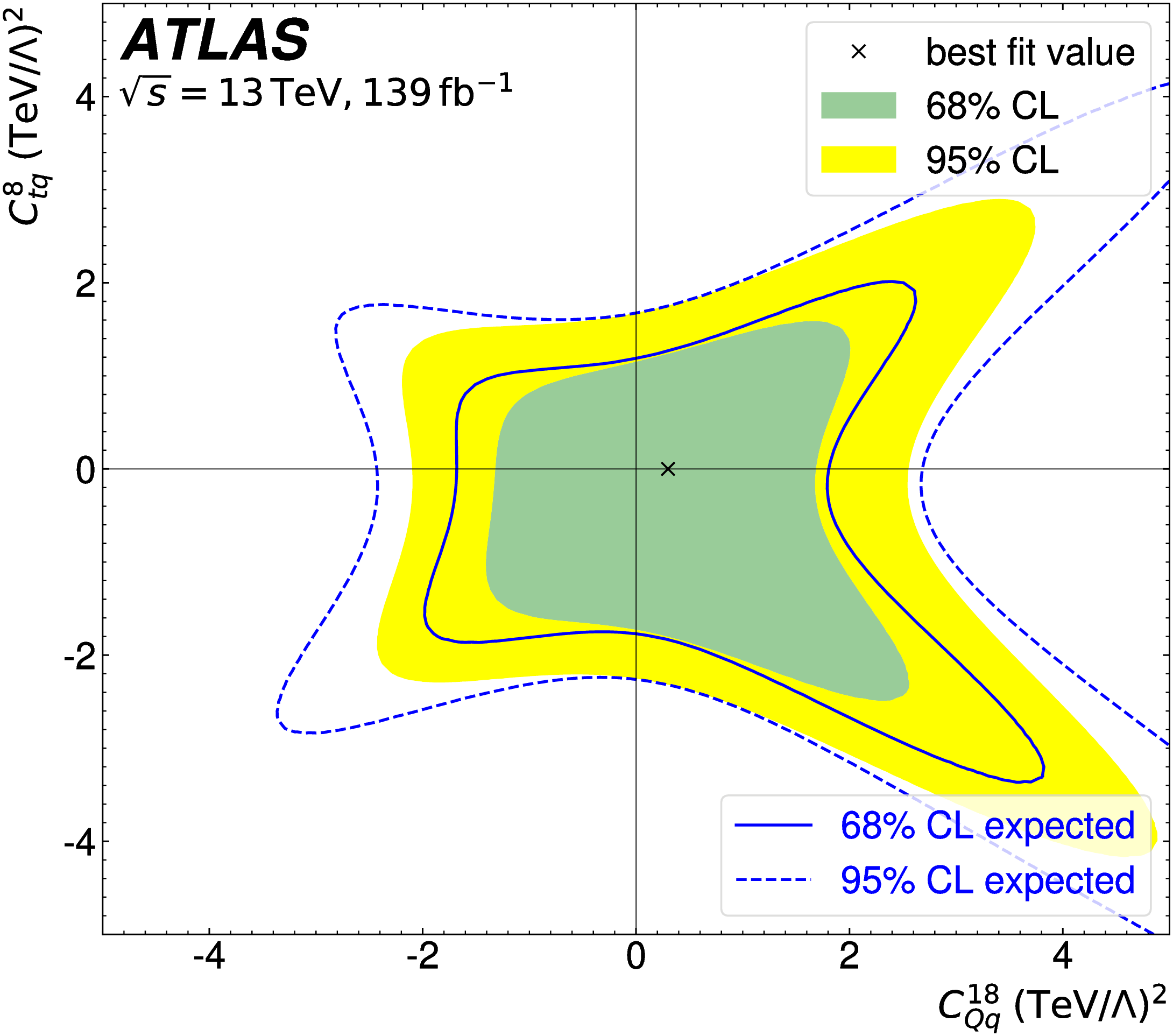}}
	\end{minipage}
	\caption[]{Two dimensional profile of Wilson coefficients for effective couplings with a singlet (left) and octet (right) color structure with left-handed (vertical axes) and right-handed (horizontal axes) top quarks.}
	\label{fig1}
\end{figure}

The measurement of this energy asymmetry is sensitive to the chirality and color charge of the associated top (anti) quarks. 
Therefore, an additional interpretation in the EFT framework is performed, probing six different four-quark EFT operators which can affect the chirality and color charges of $t\bar{t}$ production and thereby the measured energy asymmetries.
The probed dimension-6 Wilson coefficients are all found to be compatible with zero. 
%The strictest limits are achieved in a $\Lambda^{-4}$ model where also self-interacting terms of the introduced coupling modifications are taken into account.
Two dimensional profiles of the Wilson coefficients show a higher sensitivity on operators corresponding to an effective coupling with an octet color structure, compared to operators corresponding to effective couplings with a singlet color structure.
Furthermore, color singlet operators with right and left handed top quarks cannot be constrained in a two dimensional plane where the absolute values of both operators are the same. These blind spots are reduced for color octet operators with right and left handed top quarks due to interferences with gluons.
These two dimensional profiles are shown in Figure~\ref{fig1}.

\section{Top (anti) quark polarization}

At the ATLAS experiment~\cite{polarization} a measurement of the polarization of top (anti) quarks is performed in the single top t-channel production mode.
Top quarks are expected to be polarized to a large degree along the momentum of the spectator quark in single top t-channel production.
Similarly, top anti quarks are expected to be polarized to a large degree along the momentum of the incoming light quark.
These two directions are correlated strongly due to the high rapidity expected from the spectator quark.

\begin{figure}
	\begin{minipage}{0.45\linewidth}
		\centerline{\includegraphics[width=0.99\linewidth]{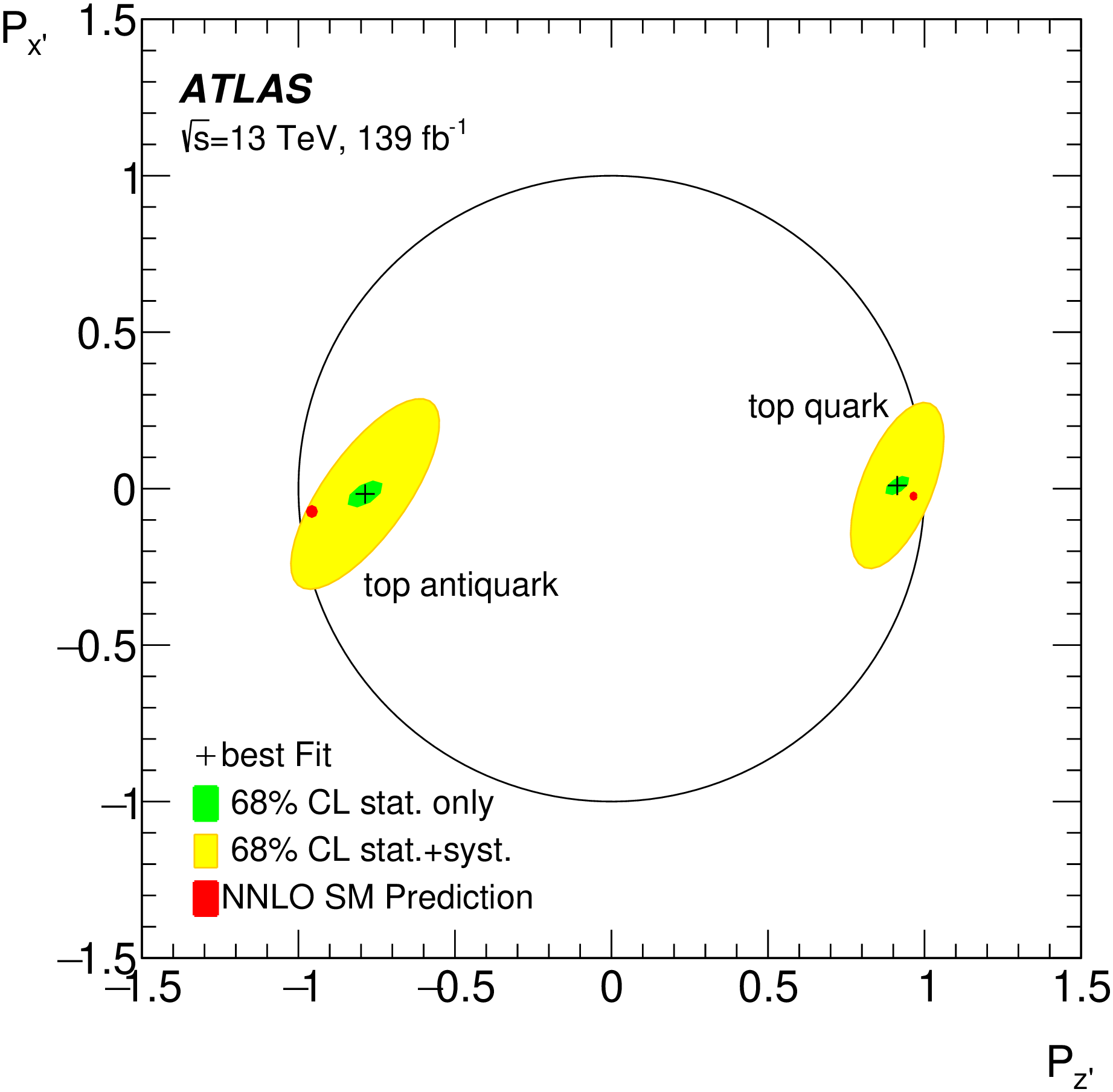}}
	\end{minipage}
	\hfill
	\begin{minipage}{0.45\linewidth}
		\centerline{\includegraphics[width=0.99\linewidth]{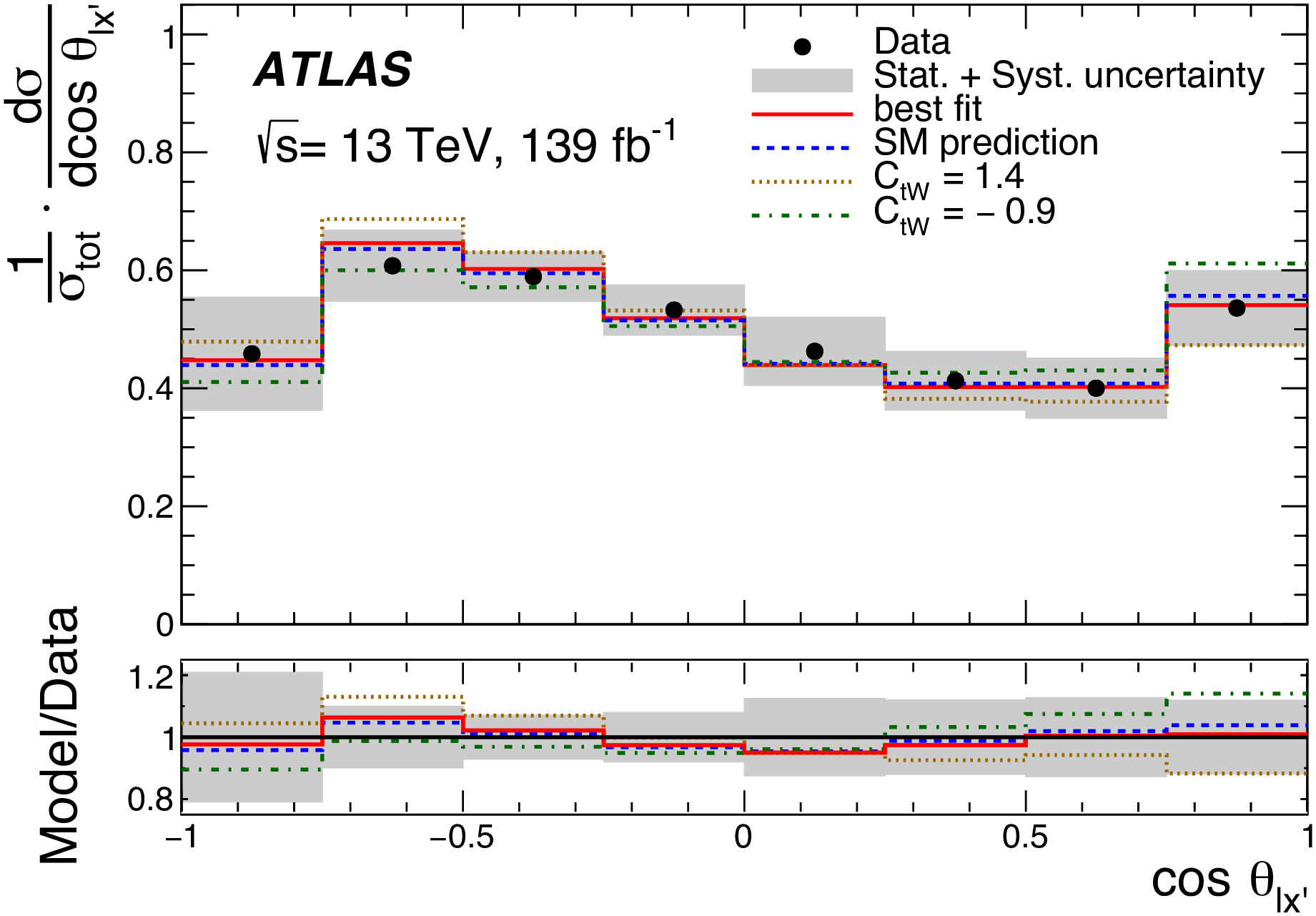}}
	\end{minipage}
	\caption[]{Left: Polarization of top quarks and anti quarks in the two dimensional $P_{z'}$ and $P_{x'}$ plane. The $z'$ direction is defined along the spectator quark momentum in the top (anti) quark rest frame. Right: Differential distribution of the charged lepton polarization angle in the $x'$ direction. Dashed and dotted lines show different hypotheses for non-zero Wilson coefficients.}
	\label{fig2}
\end{figure}

Events are categorized into eight octants of a coordinate system based on the direction of the charged lepton from top (anti) quark decay in the top (anti) quark reference frame. 
One axis is parameterized along the spectator quark momentum and another axis orthogonal to the incoming light quark.
The measurement yields a $+91\pm10\%$ and a $-79\pm16\%$ polarization of top quarks and anti quarks, respectively, along the spectator quark momentum. 
The polarization along the other coordinate axes are found to be compatible with zero. This result is shown in Figure~\ref{fig2} on the left. 
This measurement constitutes for the first time a separate measurement of top quark and anti quark polarization in single top quark production.

This measurement of the top (anti) quark polarization is sensitive to changes in the $tWb$ vertex present in the single top quark production mode.
Therefore, an interpretation of this measurement in the EFT framework is performed, probing the real and imaginary components of the $c_{tW}$ dimension-6 Wilson coefficient which effectively modifies the $tWb$ interaction.
A differential measurement of two of the three polarization angles is performed as shown in Figure~\ref{fig2} on the right. The polarization angles are the angles of the charged lepton w.r.t.\ the three previously defined coordinate axes in the top (anti) quark rest frame.
The resulting two dimensional profile of the Wilson coefficient components is found to be compatible with zero.

\section{Anomalous top couplings in $t\bar{t}$}

At the ATLAS experiment~\cite{ttsl} a differential measurement of boosted $t\bar{t}$ events is performed in the $1\ell$ channel. 
Anomalous top couplings in the $t\bar{t}$ production mode are probed via two dimension-6 Wilson coefficients, one representing an anomalous top-gluon interaction and the other a four-fermion interaction involving the two top (anti) quarks.
The differential distribution of the transverse momentum of the hadronically decaying top quark, is used to probe these Wilson coefficients in one and two dimensional profiles.
The results are compatible with the SM expectation of zero for both Wilson coefficients.

\begin{figure}
	\centerline{\includegraphics[width=0.45\linewidth]{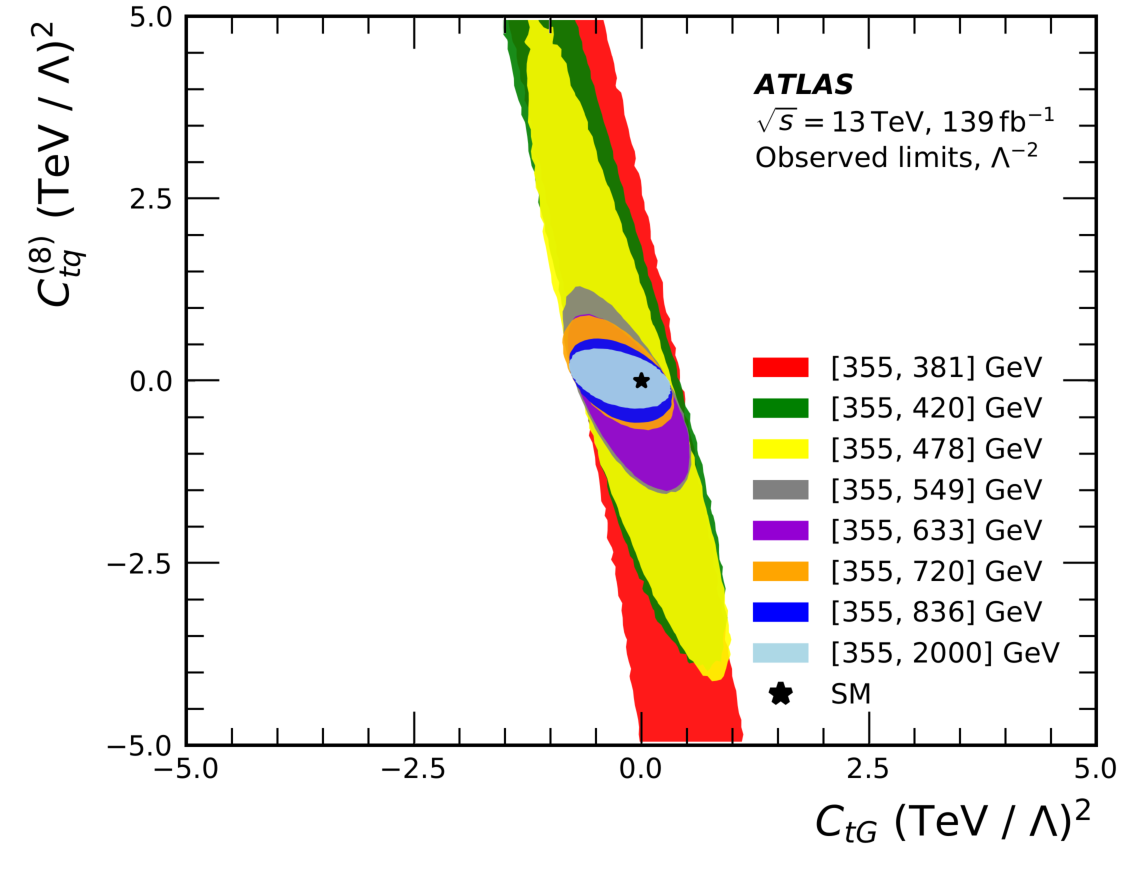}}
	\caption[]{Exclusion limits in a two dimensional profile of a top-gluon operator (horizontal axis) and a four-fermion interaction operator (horizontal axis). Different colors are different ranges of the differential distribution considered in the calculation of the limits.}
	\label{fig3}
\end{figure}

Additionally, a result shown in Figure~\ref{fig3} is presented showing the gain in sensitivity on the Wilson coefficients from differential information.
This result shows, that the sensitivity on the top-gluon interaction operator is already high when only including a single bin of the differential distribution, as this operator mostly only affects the overall $t\bar{t}$ production rate.
With only one bin included, the measurement is, however, blind to the probed four-fermion operator. 
Only after including more differential information in the measurement a similar sensitivity is achieved also on the four-fermion operator, as this operator mainly changes the differential distribution and not the overall event yield. 

Similarly, the ATLAS collaboration~\cite{ttfh} also performs a differential analysis of boosted $t\bar{t}$ events in the $0\ell$ channel.
This analysis also includes a probe of seven dimension-6 Wilson coefficients associated to four-fermion operators.
The differential distribution of the transverse momentum of the leading top quark is used in the EFT reinterpretation.
The measurement is performed in one and two dimensional profiles of the Wilson coefficients which are all compatible with zero.

\section{Anomalous top couplings in $t\bar{t}{+}\gamma$}

At the CMS collaboration differential measurements of the $t\bar{t}{+}\gamma$ process, i.e.\ the production of a prompt photon $\gamma$ in association with the $t\bar{t}$ pair, are performed. 
Two measurements are performed, in the $2\ell$~\cite{ttgammadilep} and $1\ell$~\cite{ttgammasl} channels.
In both measurements the distribution of the photon transverse momentum is reinterpreted in the EFT framework, probing the real and imaginary components of the dimension-6 $O_{tZ}$ operator.
This operator modifies the couplings of top (anti) quarks to Z bosons and photons.
One and two dimensional probes of these Wilson coefficients are performed. The results are compatible with zero and also show good agreement between the $1\ell$ and $2\ell$ channels.

\section{Anomalous top couplings in $t\bar{t}{+}Z/H$}
A measurement of $t\bar{t}{+}H$ and $t\bar{t}{+}Z$ processes is performed by the CMS collaboration~\cite{ttZH} in a boosted regime in the $1\ell$ channel.
In the measurement the soft-drop mass of a large-radius jet, which contains the collimated heavy boson decay products, is used to differentiate between Higgs an Z boson associated processes.
Additionally, a neural network is employed to differentiate between the signal processes and other $t\bar{t}$ processes.
These distributions in bins of the soft-drop mass, the neural network score and also the transverse momentum of the large-radius jet are used to determine the signal contributions of the $t\bar{t}{+}H$ and $t\bar{t}{+}Z$ processes.

As these processes give access to possible modifications in the coupling of the Z or Higgs boson to top (anti) quarks,
the above mentioned distributions are reinterpreted in the EFT framework to constrain eight dimension-6 Wilson coefficients affecting $t\bar{t}{+}Z/H$ production.
The measurement is performed independently for each Wilson coefficient and in two dimensional profiles.
All results are compatible with the SM expectations of zero.

\section{Anomalous top couplings in $t(\bar{t}){+}Z$}
Anomalous top couplings are constrained in a measurement by the CMS collaboration~\cite{tZEFT} directly designed towards a search for new top quark interactions.
The measurement targets the production of top (anti) quarks in association with Z bosons, either in $t{+}Z$ or $t\bar{t}{+}Z$ production modes.
A new measurement approach is employed, where neural networks are directly trained to identify events with anomalous top quark interactions.
Sensitivity of the neural network on events with anomalous top quark interactions is achieved by training simulated events that are SM-like against simulated events with non-zero Wilson coefficients which affect the event kinematics.
Five different dimension-6 Wilson coefficients are probed. 
The simulated events used for the training of the neural networks are sampled from a wide range of values of these five Wilson coefficients to assert sensitivity on a wide range of possible scenarios.
Results are provided for one, two and five dimensional profiles of these coefficients.
All results are compatible with the SM expectations of zero.

\section*{Acknowledgments}

The author acknowledges the support of the Doctoral School ``Karlsruhe School of Elementary and Astroparticle Physics: Science and Technology''.

\end{document}